\documentclass[a4paper]{jpconf}
\usepackage{graphicx}
\begin{document}
\title{Gluon saturation at higher orders
 and improvement of kinematics}

\author{Guillaume Beuf}

\address{Departamento de F\'isica de Part\'iculas and IGFAE, Universidade de Santiago de
Compostela,\\
 E-15706 Santiago de Compostela, Spain}

\ead{guillaume.beuf@usc.es}

\begin{abstract}
Standard perturbative calculations lead to pathologically large NLO corrections to low-$x_{Bj}$ evolution equations like BFKL and BK. Using a more refined treatment of kinematics in mixed-space, relevant when gluon saturation sets on, one obtains an improved version of the BK equation, resumming to all orders the most severe of those large higher order corrections.
\end{abstract}

\section{Introduction}

Hadronic collisions at very high energy are involving partons with very small momentum fraction in the hadronic wave-functions. Due to the high occupancy of those wee partons (mostly gluons), the phenomenon of gluon saturation 
occurs: multiple parton scattering is typical, and accompanied by strong color coherence effects. Hence, the collinear factorization (and other standard perturbative QCD formalisms) which involves only one parton from each colliding hadron does not capture the typical physics of collisions in the high-energy limit, with only a semi-hard momentum transfer.

Instead, the coherent multiple scattering effects are taken into account straightforwardly when describing the wee gluons inside each highly boosted hadron as a semi-classical gluon field (see Ref. \cite{CGC_rev} and references therein). The main formalism based on this idea is the Color-Glass-Condensate effective theory (CGC). Each ultra-relativistic nucleus is described by a random classical shockwave gluon field with a classical statistical distribution, 
 and QCD quantum corrections are resummed within leading logarithmic accuracy (LL) at small momentum fraction by the JIMWLK evolution 
 of the gluon field distribution. The JIMWLK functional equation can also be written as Balitsky's infinite hierarchy of equations. 

Inclusive enough observables, like DIS structure functions at low $x_{Bj}$ or single inclusive particle production at forward rapidity in pA collisions, can be expressed in terms of the scattering amplitude of a color dipole on the gluon field of the target. For that object, the Balitsky-Kovchegov (BK) equation \cite{Balitsky:1995ub,Kovchegov:1999yj,Kovchegov:1999ua} gives a safe approximation of the full JIMWLK equation. Adding running coupling effects to the BK equation \cite{Balitsky:2006wa,Kovchegov:2006vj} leads to a successful phenomenological description of DIS data at HERA \cite{Albacete:2010sy,Kuokkanen:2011je} within the CGC, as well as of forward particle production at RHIC \cite{Albacete:2010bs}.

In the recent years, the calculation of NLO corrections in that framework has been a hot topic. After the running coupling corrections \cite{Balitsky:2006wa,Kovchegov:2006vj}, the full set of NLO corrections to the BK equation have been calculated \cite{Balitsky:2008zza}. Later, the calculation of the NLO corrections has been performed for the \emph{impact factor} or \emph{coefficient function} both in the case of DIS structure functions \cite{Balitsky:2010ze,Beuf:2011xd} and of forward hadron production in pA collisions \cite{Chirilli:2012jd}.

Unfortunately, those NLO calculations cannot be used, in the form in which they are now available, to perform phenomenological studies at full NLO accuracy. Indeed, the BK equation at NLO suffers from the same problem as its linear version, the BFKL 
equation at NLO \cite{Fadin:1998py,Ciafaloni:1998gs}: some of the NLO corrections are pathologically large and lead an instability of the solutions. This signals a breakdown of the perturbative expansion as done usually in the Regge limit. Those large NLO corrections are due to the inability of the standard perturbative expansion in the Regge limit to provide results matching smoothly with DGLAP physics in the collinear and in the anticollinear regimes \cite{Salam:1998tj}. Hence, the large higher order corrections to the BFKL and BK equations can be resummed to all orders by performing an appropriate matching with the DGLAP equation at LO (or beyond) in the collinear and in the anticollinear regimes \cite{Salam:1998tj}. That program has been completed for the BFKL equation both in momentum space \cite{Ciafaloni:2003rd} and in Mellin space \cite{Altarelli:2005ni}. However, the BK equation is more naturally written in mixed space. The generalization of that resummation to the case of the BK equation requires a significant effort mostly due to the translation to mixed space, and to a lesser extent due to the nonlinearity of the BK equation.

Among the large higher order corrections to be resummed, the ones of purely kinematical origin are the most severe, but also the easiest to deal with \cite{Salam:1998tj}. In section \ref{sec:DiagnosingIssues} of this contribution, the NLO impact factors for DIS \cite{Beuf:2011xd} are analysed to understand why such kinematical issues arise. And in the section \ref{sec:kcBK}, an improved version of the BK equation at LO is proposed, which realizes the resummation of those large kinematical higher order corrections. It corresponds to the mixed space\footnote{In mixed space, the kinematics of partons is described by their light-cone momentum $k^+$ and their transverse position $\mathbf{x}$.} analog of the kinematical constraint \cite{Ciafaloni:1987ur,Andersson:1995ju,Kwiecinski:1996td} in momentum space. It also represents a first step towards a full resummation providing a fully stable and reliable version of the BK equation at NLO.

\section{Diagnosing kinematical issues from the explicit NLO impact factors for DIS\label{sec:DiagnosingIssues}}

The DIS structure functions are linear combinations of the total cross sections for the scattering of a transverse or longitudinal virtual photon off the target, which at strict NLO accuracy in the CGC can be written as \cite{Beuf:2011xd} (see also \cite{Balitsky:2010ze})
\begin{eqnarray}
 \sigma_{T,L}^{\gamma} &=&  2\; \frac{2N_c\, \alpha_{em}}{(2\pi)^2}\sum_f e_f^2  \int \textrm{d}^2\mathbf{x}_{0} \int \textrm{d}^2\mathbf{x}_{1} \int_0^1 \textrm{d} z_1\,  \Bigg\{ \mathcal{I}_{T,L}^{LO}({x}_{01},z_1)\; \Big[1-  \left\langle {\cal S}_{01} \right\rangle_0\Big] \nonumber\\
& &    +  \frac{N_c\, \alpha_s}{\pi}  \int \frac{\textrm{d}^2\mathbf{x}_{2}}{2\pi} \int_{0}^{1\!-\!z_1}\frac{\textrm{d}z_2}{z_2}\; \mathcal{I}_{T,L}^{NLO}(\mathbf{x}_{0},\mathbf{x}_{1},\mathbf{x}_{2},z_1,z_2)\;
\left\langle {\cal S}_{01}- {\cal S}_{02}\, {\cal S}_{21}\right\rangle_0
\Bigg\}\, ,\label{DipoleFact_strictNLO}
\end{eqnarray}
where ${\cal S}_{ij}$ is the S-matrix for the scattering of a fundamental color dipole with transverse positions $\mathbf{x}_{i}$ and $\mathbf{x}_{j}$ off a gluon shockwave, $\langle \dots \rangle_0$ is the statistical average over the target's gluon field with no LL quantum corrections included, and ${x}_{ij}=|\mathbf{x}_{i}-\mathbf{x}_{j}|$. The LO impact factors $\mathcal{I}_{T,L}^{LO}$ have been known for a long time \cite{Bjorken:1970ah,Nikolaev:1990ja}, whereas the NLO ones $\mathcal{I}_{T,L}^{NLO}$ have been calculated in Refs. \cite{Balitsky:2010ze,Beuf:2011xd}. The integral over the photon's momentum fraction $z_2=k_2^+/q^+$ carried by the radiated gluon is logarithmically divergent for $z_2\rightarrow 0$, but in that limit
\begin{equation}
\mathcal{I}_{T,L}^{NLO}(\mathbf{x}_{0},\mathbf{x}_{1},\mathbf{x}_{2},z_1,z_2=0) = \frac{x_{01}^2}{x_{02}^2\, x_{21}^2}\; \mathcal{I}_{T,L}^{LO}({x}_{01},z_1)\, .\label{Fact_IF_NLO_low_z2}
\end{equation}
Together with an appropriate factorization scheme (including for example a cut-off in $k^+$), the BK equation
\begin{equation}
\partial_{Y^+}  \left\langle {\cal S}_{01} \right\rangle_{Y^+}=   \frac{N_c\, \alpha_s}{\pi}
\int  \frac{\textrm{d}^2\mathbf{x}_{2}}{2\pi} \frac{x_{01}^2}{x_{02}^2\, x_{21}^2}\:
\left\langle{\cal S}_{02} {\cal S}_{21}\!-\! {\cal S}_{01}\right\rangle_{Y^+}\label{BK_LL}
\end{equation}
allows to resum those small $z_2$ LL contributions. In that case, one should use in the first line of the expression (\ref{DipoleFact_strictNLO}) the dipole S-matrix $\left\langle {\cal S}_{01} \right\rangle_{Y_f^+}$ evolved with the BK equation (\ref{BK_LL}) over a range $Y_f^+=\log(k_f^+/k_{\min}^+)$. $k_{\min}^+$ is the typical $k^+$ scale set by the target and $k_f^+$ an appropriate factorization scale in $k^+$, such as $k_f^+=z_1(1\!-\!z_1)q^+$.

The LO (resp. NLO) impact factor $\mathcal{I}_{T,L}^{LO}$ (resp. $\mathcal{I}_{T,L}^{NLO}$) contains a factor which suppresses exponentially the large values of $Q^2\, X_2^2$ (resp. $Q^2\, X_3^2$), where
\begin{equation}
X_2^2= z_1\, (1\!-\!z_1)\, {x}_{01}^2 \quad \textrm{and} \quad X_3^2= z_1\, (1\!-\!z_1\!-\!z_2)\, {x}_{01}^2 + z_2\, (1\!-\!z_1\!-\!z_2)\, {x}_{02}^2 + z_2\, z_1\, {x}_{21}^2\, .
\end{equation}
As argued in Ref. \cite{Beuf:2011xd} the variables $Q^2\, X_2^2$ and $Q^2\, X_3^2$ are the ratios of the formation time of the quark-antiquark or quark-antiquark-gluon Fock components of the photon, resolved by interaction with the target, over the lifetime of the virtual photon. Hence, the interpretation of that exponential suppression is very clear: a Fock state which has not enough time to be formed as fluctuation of the virtual photon within the lifetime of the latter cannot give a non-negligible  contribution to the DIS cross sections.

The standard treatment of low $z_2$ LL with the BK equation discussed previously requires to approximate $X_3^2$ by $X_2^2$, in order to obtain the factorization (\ref{Fact_IF_NLO_low_z2}) of $\mathcal{I}_{T,L}^{NLO}$. Although exact at $z_2=0$, the approximation $X_3^2\simeq X_2^2$ is not generically correct at small but finite $z_2$: it is wrong when the gluon is emitted at a so distant transverse position $\mathbf{x}_{2}$ that $z_1(1\!-\!z_1)x_{01}^2 \ll z_2 x_{02}^2 \simeq z_2 x_{12}^2$. In that regime, not only the nice feature of suppression of Fock states too long to form is spoiled by the standard subtraction of LL, but also the term used to subtract the LL contributions from the NLO term in the expression (\ref{DipoleFact_strictNLO}) is parametrically larger than both the unsubtracted NLO term and the LO term, which signals a breakdown of this formalism. 

Evolution equations like BK and BFKL can be derived from the knowledge of the photon impact factor at arbitrary order but restricted to the case of softer and softer gluons emitted successively \cite{Mueller:1993rr,Kovchegov:1999yj}. Usually, all the transverse scales are assumed to be of the same order in that context. This assumption (also used in other derivations of those equations) is not completely self-consistent due to the unrestricted integration over transverse momentum or position in the evolution kernel. The resulting issue is essentially the same as found in the study of the NLO photon impact factor: in the parton cascades resummed by the LO BFKL and BK equations, the softer and softer gluons are not always correctly ordered in formation time. The most pathological higher order corrections to the BFKL and BK kernel are then induced by that little inconsistency at LO.

\section{Improving the treatment of kinematics in the BK equation\label{sec:kcBK}}

According to our previous discussion, the standard BK equation at LO (\ref{BK_LL}) includes, at large $\mathbf{x}_{2}$, unphysical contributions from gluons which should not have time to be formed. The first step to cure this problem is to modify the probability density of soft real gluon emission by a color dipole by forbidding emissions with $z_1(1\!-\!z_1)x_{01}^2 \ll z_2 x_{02}^2 \simeq z_2 x_{12}^2$. This is the mixed-space analog of the kinematical (\emph{a.k.a.} consistency) constraint of Refs. \cite{Ciafaloni:1987ur,Andersson:1995ju,Kwiecinski:1996td}.
In mixed-space, this restriction has been first proposed in Ref. \cite{Motyka:2009gi}, where it was inferred from the structure of $n$-gluons MHV amplitudes. The resummation scheme proposed here differs however from the one in Ref. \cite{Motyka:2009gi} in several aspects, most notably in the treatment of virtual corrections.

The choice done in Ref. \cite{me} is to organize the virtual corrections in such a way that the probabilistic interpretation of the dipole cascade \cite{Mueller:1993rr} is maintained at each order once the resummation is done. Then, the modified virtual corrections can be calculated unambiguously from the modified real gluon emission probability. Performing this task and writing the result as an evolution equation, one gets the improved BK equation with kinematical constraint \cite{me}
\begin{eqnarray}
& &\!\!\!\!\!\!\!\!\!\! \partial_{Y^+}  \left\langle {\cal S}_{01} \right\rangle_{Y^+}=   \frac{N_c\, \alpha_s}{\pi}
\int  \frac{\textrm{d}^2\mathbf{x}_{2}}{2\pi} \frac{x_{01}^2}{x_{02}^2\, x_{21}^2}\: \theta(Y^+\!-\!\Delta_{012})\; \Bigg\{
\left\langle{\cal S}_{02} {\cal S}_{21}\!-\! \frac{1}{N_c^2} {\cal S}_{01}\right\rangle_{Y^+\!-\!\Delta_{012}}
 \nonumber\\
& &
\qquad \qquad \qquad \qquad \qquad \qquad \qquad \qquad \qquad \qquad \qquad -\! \left(1\!-\! \frac{1}{N_c^2}\right) \left\langle{\cal S}_{01} \right\rangle_{Y^+} \Bigg\}\, .
\label{kcBK}
\end{eqnarray}
A convenient definition for the shift $\Delta_{012}$ (but not unique, due to some resummation scheme ambiguity) is
\begin{equation}
\Delta_{012}= \max \left\{0,\, \log\left(\frac{\min(x_{02}^2, x_{21}^2)}{x_{01}^2}\right)  \right\}\label{Delta_shift}\, .
\end{equation}
Due to the theta function in (\ref{kcBK}), the phase space for gluon emission at large $\mathbf{x}_{2}$ is severely restricted at small $Y^+$, but progressively opens up in the course of the $Y^+$ evolution. The shift of $Y^+$ in the real term only also contributes to slow down the evolution with respect to the standard BK equation (\ref{BK_LL}).

The largest NLO corrections \cite{Balitsky:2008zza} to the BK equation are indeed resummed into the improved LO equation (\ref{kcBK}). However, a full resummation of all pathological NLO corrections as in Refs. \cite{Ciafaloni:2003rd,Altarelli:2005ni} requires further work.
The equation (\ref{kcBK}) also allows to subtract the LL contributions from the NLO impact factors (\ref{DipoleFact_strictNLO}) in a correct way in all the phase-space, by contrast to the equation (\ref{BK_LL}).

Implementing the kinematical improvement (\ref{kcBK}) together with running coupling corrections \cite{Balitsky:2006wa} should lead to solutions more naturally in good phenomenological agreement with the DIS data, due to a slower $Y^+$ evolution \cite{Albacete:2010sy,Kuokkanen:2011je}.

\end{document}